\documentclass[twocolumn,nofootinbib,superscriptaddress]{revtex4}
\usepackage{amsfonts,amsmath,mathrsfs,epsfig,amsbsy,bm,verbatim}

\usepackage[usenames,dvipsnames]{xcolor}

\usepackage[pdftex, bookmarks={false},
pdfauthor={Brian Swingle, John McGreevy}, pdftitle={s-sourcery}]{hyperref}
\hypersetup{colorlinks=true, linkcolor=BrickRed, citecolor=Violet, filecolor=OliveGreen, urlcolor=RoyalBlue, filebordercolor={.8 .8 1}, urlbordercolor={.8 .8 0}}

\usepackage{graphicx}

\pdfoutput=1

\newcommand{\beq}{\begin{equation}}
\newcommand{\eeq}{\end{equation}}
\newcommand{\bea}{\begin{eqnarray}}
\newcommand{\eea}{\end{eqnarray}}

\newcommand{\tr}{\text{tr}}

\begin{document}
\title{Area Law for Gapless States from Local Entanglement Thermodynamics}
\author{Brian Swingle}
\affiliation{Department of Physics, Stanford University, Stanford, CA 94305, USA}

\author{John McGreevy}
\affiliation{Department of Physics,
University of California at San Diego,
La Jolla, CA 92093, USA}

\begin{abstract}
We demonstrate an area law bound on the ground state entanglement entropy of a wide class of
gapless quantum states of matter using a strategy called
local entanglement thermodynamics. The bound depends only on thermodynamic data, actually a single exponent, the hyper-scaling violation exponent $\theta$.  All systems in $d$ spatial dimensions obeying our scaling assumptions and with $\theta < d-1$ obey the area law, while systems with $\theta = d-1$ can violate the area law at most logarithmically. We also discuss the case of frustration-free Hamiltonians and show that to violate the area law more than logarithmically these systems must have an unusually large number of low energy states. Finally, we make contact with the recently proposed $s$-source framework and argue that $\theta$ and $s$ are related by $s=2^\theta$.
\end{abstract}

\maketitle

When a quantum many-body system is at or near zero temperature, quantum entanglement between the constituents leads to qualitatively new phenomena (for a review and references, see {\it e.g.}~\cite{2012arXiv1210.1281W, 2013NJPh...15b5002G}).
Although direct measurement of many-body entanglement awaits advances in quantum state engineering,
it is already an invaluable theoretical tool for diagnosing the physics of quantum many-body systems.
An important role is played by the entanglement entropy
of a spatial region, which quantifies the amount of bipartite entanglement between the region and its complement.

Here we demonstrate that the ground state entanglement entropy obeys an area law for a wide variety of gapless and scale invariant phases of matter. While the area law is widely believed to hold for gapped phases ({\it e.g.}~\cite{Eisert:2008ur}),
certain gapless phases are definite exceptions, notably conventional metals \cite{PhysRevLett.96.010404, PhysRevLett.96.100503, Swingle:2009bf, Swingle:2010yi}
and conformal field theories (CFTs) in $1+1$ dimensions \cite{Calabrese:2009qy}.
In addition to demonstrating an area law, our results expose a deep connection between thermodynamic properties and entanglement properties in scale invariant phases of matter. Such a connection brings entanglement closer to experimentally accessible probes.
Previous work relating the scaling of geometric entanglement
to thermodynamics and energy fluctuations includes
\cite{2014arXiv1409.5946B, 2009PhRvA..80e2104M}.

Given a bi-partite quantum state $|\psi_{A\bar{A}}\rangle$ we study the
reduced state of the $A$ subsystem, $\rho_A = \text{tr}_{\bar{A}}(|\psi_{A\bar{A}}\rangle \langle \psi_{A\bar{A}} |)$, and in particular its entanglement entropy, $S(A) = - \text{tr}_A(\rho_A \log(\rho_A))$. When $|\psi_{A\bar{A}}\rangle$ is the ground state of a local Hamiltonian the entanglement entropy often obeys an area law, $S(A) \sim |\partial A|$ meaning the entropy is proportional to the size of the boundary of $A$. An area law's worth of entanglement always appears due to short-distance correlations, but to have more than an area law's worth of entanglement intuitively requires some long-range structure in the quantum state. For example, conventional metals with their Fermi surface of low energy electronic excitations violate the area law with a multiplicative logarithmic correction~\cite{PhysRevLett.96.010404, PhysRevLett.96.100503, Swingle:2009bf, Swingle:2010yi}. The intuition for this violation is that metals have a great many spatially extended low energy excitations and this plethora of low energy modes leads to a plentitude of long-range entanglement in the ground state. It is the purpose of this paper to make this intuition more precise and to extend it to other kinds of scale invariant states of matter.

The argument appeals to ordinary thermodynamic properties of the state of matter to quantify the number of low energy excitations. The connection between thermodynamics and entanglement then proceeds by recasting the entanglement entropy problem as a problem of local thermodynamics, that is, thermodynamics with a locally varying temperature. In fact, given a state $\rho_A$ arising from a scale invariant ground state, there is another state $\sigma_A$ of a local thermodynamic form, whose entropy bounds that of $\rho_A$.

Assuming the thermal entropy scales with temperature like $s(T) \sim T^{\frac{d-\theta}{z}}$ with $\theta$ the hyperscaling violation exponent and $z$ the dynamical exponent, we show that the corresponding ground state obeys the area law provided $\theta < d-1$
and $z$ is positive and finite. The case $\theta = d-1$ (which occurs for conventional metals) is marginal and leads to a multiplicative logarithmic violation to the area law. Physically the only assumptions are translation invariance, scale invariance, and the validity of local thermodynamics within a derivative expansion; the precise assumptions are stated just below.

\textit{Problem setup.} Consider the ground state $|g\rangle = |\psi_{A\bar{A}}\rangle$ of a local $d$-dimensional Hamiltonian $H = \sum_x H_x$ defined on $L^d$ sites (energy scale $J$, range $\ell$) with a $1/\text{poly}(L)$ gap which supports scale invariant physics. Let $\rho_A = \tr_{\bar{A}}(|g\rangle \langle g|)$ be the reduced density matrix of region $A$, and let $\sigma_A$ denote the maximum entropy state consistent with the expectation values of all the $H_x$ contained in $A$. In other words, $\sigma_A$ is the state of maximum entropy which gives the same expectation values as $\rho_A$ for all the terms in the Hamiltonian contained in region $A$.

Since $\rho_A$ is consistent with its own local data, it follows that $S(\sigma_A)\geq S(\rho_A)$. Furthermore, $\sigma_A$ is a local Gibbs state:
\beq
\sigma_A = \frac{\exp\left(-  \sum_{x\in A} H_x/T(x) \right)}{Z}
\eeq
where again the $H_x$ denote the terms in the local Hamiltonian and the $T(x)$ are constants adjusted so that $\text{tr}(\rho_A H_x) = \text{tr}(\sigma_A H_x)$. This local Gibbs form arises from maximizing the entropy subject to the constraint that local data is correctly reproduced \cite{2010NatCo...1E.149C, 2014arXiv1407.2658S}.

Let $H_A$ denote the restriction of $H$ to region $A$. $H_A$ is a sum of local terms residing within $A$, $H_A = \sum_{x\in A} H_x$. The locality of $H$ implies that $\tr(H_A \rho_A)$ is within $J |\partial A| \ell$ of the ground state energy of $H_A$ (where $J \equiv ||H_x||$ measures the coupling strength).
The bound arises because the lost terms in $H$ are localized at the boundary $\partial A$, so $\rho_A$ is a ground state of $H_A$ up to excitations near the boundary.

As a local Gibbs state, $\sigma_A$ is essentially a thermal state of $H_A$ with a position dependent temperature $T(x)$. The form of $T(x)$ is tightly constrained since the expectation value of $H_x$ is independent of $x$ in the ground state and thus in the state $\sigma_A$ as well. Furthermore, the local temperature approaches zero away from the boundary to ensure that the average excitation energy relative to the ground state of $H_A$ is localized near the boundary of $A$. Hence a picture emerges wherein the entropy of $\sigma_A$ is concentrated near $\partial A$. We now make this idea sharp.

The crucial observation is that we can estimate the scaling of the energy and entropy of $\sigma_A$ using local (intensive) thermodynamic expressions. In other words, if $e(T)$ and $s(T)$ are the bulk thermodynamic energy and entropy densities respectively, then
for purposes of studying the scaling with region size, we may make the replacements
\beq
\tr(H_A \sigma_A) \sim E_{g,A}+\int d^d x\, e(T(x)),
\eeq
and
\beq
- \tr(\sigma_A \log(\sigma_A)) \sim \int d^d x\, s(T(x)),
\eeq
where $T(x)$ is the local temperature (inverse of the local coefficient of $H_x$ in $-\log(\sigma_A)$). As far as bounding the entropy is concerned,
local thermodynamics is a good approximation because neglecting correlations between distant regions (that is, treating the system locally) should only increase the effective entropy. Later we give a more careful justification of the assumption of local thermodynamics.

To use this assumption we must specify the thermodynamic properties of the scale invariant phase. The relevant thermodynamic scaling data are the dynamical exponent $z$ and the hyper-scaling violation exponent $\theta$. $z$ relates energy to momentum as $\omega \sim k^z$. $\theta$ controls the scaling of the entropy density with length: $T^{-\frac{1}{z}}$ is a length and $s$ is assumed to scale as $d-\theta$ powers of inverse length, so $s(T) \sim T^{\frac{d-\theta}{z}}$. $\theta$ controls the extent to which the naive scaling with density fails for the entropy (and other thermodynamic quantities), hence its name. In systems where $\theta$ is non-zero, other fixed microscopic length scales make up the units of the entropy density. In the example of a Fermi surface, the units are made up by $\theta=d-1$ powers of the Fermi momentum. The scaling of the energy density is determined by thermodynamics to be $e(T) \sim T s(T)$.

An important aspect of the resulting theory of {\it entanglement thermodynamics}
is that it is characterized by the scaling exponents $z, \theta$
of the fixed point $H$ itself. In particular, the effective temperature $T(x)$ must vary smoothly with $x$ in a way controlled by $z$ and $\theta$ to give a translation invariant expectation value for $H_x$ (although oscillations with other fixed length scales when $\theta \neq 0$ are not ruled out). Remember also that if the subregion $A$ grows to encompass the total system then the exact ground state is recovered \cite{2010NatCo...1E.149C, 2014arXiv1407.2658S}, so it must be true that $T(x)$ decays as $x$ moves away from the boundary of $A$. For future reference, note that for frustration-free Hamiltonians the above analysis further simplifies since $T(x) = 0$ identically.

\textit{Entanglement entropy bound.} Now we estimate the entropy in a simple geometry. Suppose $A$ is a half-space in $d$ dimensions with translation invariance in $d-1$ transverse dimensions. We compactify these transverse directions to have size $R$.  Translation invariance implies that $T(x)$ depends only on $x$, the distance from the boundary.  We also introduce a short distance cut-off, $a$, which could be the lattice spacing, and a long-distance cutoff, $w$, which could be the width of the half-space (making it actually a long strip of length $R \gg w$).
\includegraphics[width=.5\textwidth]{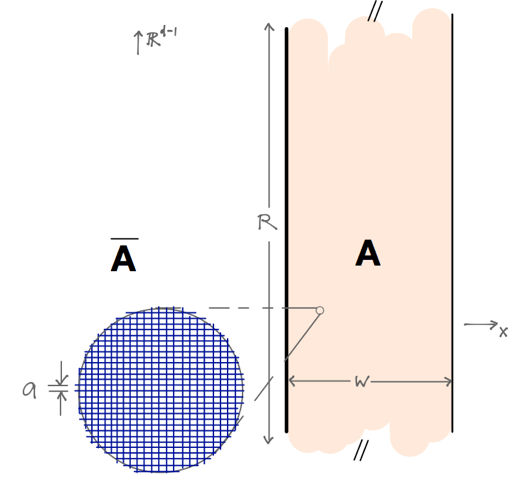}
The scaling of energy with momentum (inverse length) and the absence of any other scale determines the local temperature $T(x)$ to be
\beq
T(x) \sim x^{-z}.
\eeq
In fact, two other solutions consistent with scale invariance are $T=0$ and $T=\infty$, and while $T=\infty$ can be ruled out (too much energy), $T=0$ is actually relevant in the context of frustration-free Hamiltonians as discussed below.
For the special case of Lorentz-invariant systems,
this relation is rigorous, since
the entanglement Hamiltonian for a half space
($ \equiv - \log \rho_\text{half-space}$)
is a boost generator
\cite{bwtheorem, 1976PhRvD..14..870U,1973PhRvD...7.2850F,1975JPhA....8..609D,PhysRevD.29.1656, 2012PhRvB..86d5117S}.
The scaling form for $T(x)$ is further justified in an appendix. Assuming this form for $T(x)$, the energy density is
\beq
e(T(x)) \sim x^{-z + \theta-d}
\eeq
and the entropy density is
\beq
s(T(x)) \sim x^{\theta-d}.
\eeq

Integrating the energy density gives
\beq
\tr(H_A \sigma_A)\sim E_{g,A} + R^{d-1} \int_a^w dx \,x^{-z - d + \theta}
\eeq
which converges as the long-distance cutoff $w$ is taken to infinity provided $z + d > 1 + \theta$.  For example, any relativistic CFT has $\theta=0$ and $z=1$, so the integral converges for all $d$ in this case. Since we must have $\tr(H_A \sigma_A) \leq E_{g,A} + J |\partial A| \ell$ we see that any local theory which obeys our assumptions must have $z+d > 1 + \theta$.

Turning to the entropy density, we have
\beq \label{eq:entestimate}
- \tr(\sigma_A \log(\sigma_A)) \sim R^{d-1} \int_a^w dx\,x^{- d + \theta},
\eeq
so that the integral converges provided $d > 1+\theta$, independent of the value of $z$!
Since \eqref{eq:entestimate} is an upper bound on $S_A$, all phases of matter with $\theta < d-1$ therefore obey the area law. This is our main result. The case $\theta = d-1$, realized in Fermi liquids, is marginal and gives a logarithmic divergence with $w$. Since we are upper bounding the entropy, it does not follow that phases with $\theta = d-1$ must violate the area law. To violate the area law worse than logarithmically, a phase must have $\theta > d-1$ (or have a large number of ground states).

Such an inequality was previously deduced \cite{2012PhRvB..85c5121H}
for the special case of systems with a classical gravity dual.

\textit{Validity of the local approximation.} We now give a detailed justification for the assumption of local thermodynamics. The main idea is to make the local approximation better by going to higher ``temperature."
A similar construction was used for other purposes in \cite{Grover:2011fa}.
Define $\tilde{H}_A$ by the equation $\sigma_A = \frac{e^{-\tilde{H}_A}}{Z}$ and then define
\beq
\sigma_A(\tau) = \frac{e^{-\tilde{H}_A/\tau}}{Z(\tau)}
\eeq
with
$Z(\tau)\equiv \tr e^{-\tilde{H}_A/\tau}$.
$\tau$ is a fictitious temperature such that $\tau = 1$ is the original maximum entropy state $\sigma_A$.  Ordinary thermodynamics for the local Hamiltonian $\tilde{H}_A$ implies that the entropy of $\sigma_A(\tau)$ is a monotonic function of $\tau$. Furthermore, as $\tau$ grows, the local approximation becomes better and better because the correlation length becomes shorter (modulo encountering a classical phase transition\footnote{It would be fascinating to find models where one is
{\it forced} to encounter such a classical phase transition upon raising the temperature,
and hence where local thermodynamics need not apply!}).

Returning to our entropy estimate above, we now have $T(x) \sim \tau x^{-z}$ and the local approximation gives
\beq
S(\sigma_A(\tau)) = \tau^{\frac{d-\theta}{z}} S(\sigma_A)\sim \tau^{\frac{d-\theta}{z}}R^{d-1} \int_a^w dx\,x^{- d + \theta}.
\eeq
Thus for any large but $R$- and $w$-independent $\tau$, $S(\tau)$ scales with $R$ and $w$ the same way as the local approximation for $S(1)$ and we expect smaller error from the local approximation.

To give a precise characterization of validity of the local thermodynamic approximation, we make use of a derivative expansion argument familiar from hydrodynamics. The local applicability of thermodynamics follows if all perturbations vary slowly on the scale of the correlation length: if this is so, then the system responds locally to the perturbation because the memory of distant regions is effectively washed out.  In an interacting scale invariant system, the correlation length at finite temperature scales as $\xi(T) \sim T^{-1/z}$. Note that free theories, e.g., Goldstone bosons \cite{2012PhRvL.108l0401C, 2011arXiv1112.5166M}, may not develop such a correlation length at finite temperature, but being free theories, we may independently establish the area law (or lack thereof). Since the spatially varying temperature is the only perturbation, the validity of local thermodynamics in the derivative expansion rests on the condition
\beq
\delta \equiv \xi \frac{|\nabla T|}{T} \ll 1.
\eeq

In the half-space geometry with $\tau=1$ we argued that $T(x) \sim x^{-z}$, so we have $\xi \sim x$ and the left hand side of the derivative expansion condition is of order one, $\delta \sim 1$.  Hence the derivative expansion is not obviously well-controlled in this case. However, we can improve the situation by introducing a large but system size independent $\tau \gg 1$.  Then because the derivative expansion condition is inhomogeneous in $T$ we have the modified condition
\beq
\delta(\tau) = \underbrace{\left(\frac{x}{\tau^{1/z}} \right)}_{\xi} \underbrace{\frac{z}{x}}_{\frac{|\nabla T|}{T}} = \frac{z}{\tau^{1/z}} \ll 1.
\eeq
In the last step, we assumed $ 0 < z < \infty$.
Thus within a formal derivative expansion the large $\tau$ state should obey local thermodynamics.\footnote{For an extensive quantity $f$ the derivative expansion implies a correction of the form $f = f_{local}(1 + O(e^{-1/\delta}))$ with the corrections being due to higher derivative terms which are irrelevant and thus do not change the scaling of the entropy. Paranoid readers may take $1/\delta \sim \log(R)$ to fully suppress the corrections at the expense of a weaker bound (unphysical log corrections).}

In an appendix we give two additional results related to the validity of the local approximation. The conclusion in all cases is the same, but these results give us better control over the corrections to the local approximation and may be of independent interest.

\textit{Frustration-free Hamiltonians.} Here we briefly discuss the very interesting case of frustration-free gapless Hamiltonians.  This study is motivated in part by \cite{2012PhRvL.109t7202B} which presented an interesting frustration-free Hamiltonian defined on a segment of length $L$ with the property that the gap above the unique ground state was $\frac{1}{\text{poly}(L)}$ and the entanglement entropy of one half of the segment $\sim \log(L)$. Ref.~\cite{2014arXiv1408.1657M}
solves another model that violates the area law more than logarithmically, in fact as $\sqrt{L}$.

Consider a frustration-free Hamiltonian $H$ with a unique ground state $|g\rangle$.  Recall that $H_A$ denotes the Hamiltonian restricted to $A$ and $\rho_A$ the state restricted to $A$.  By assumption, $H$ is frustration-free, so it can be written as a sum of local operators $H = \sum_x H_x$ where each $H_x$ is positive and annihilates the ground state.  Thus we have
\beq \label{zeroHA}
\tr(\rho_A H_A) = 0
\eeq
since every term in $H_A$ independently annihilates the ground state.  Hence $\rho_A$ lies within the ground state manifold of $H_A$, and as such, its entropy must be less than the maximum entropy state in the ground state, the uniform mixture of all ground states.  Hence we have $S(\rho_A) \leq \log(G(H_A))$ where $G(H_A)$ is the ground state degeneracy of $H_A$.

It is interesting to note that the maximum entropy state consistent with the local terms in $H_A$ is
the (normalized) projector onto the ground state of $H_A$,
$$ \rho_\text{max} = {1\over G(H_A)} P_{H_A=0} .$$
 Since we may write the ground state projector as
\beq
P_{H_A=0} = \lim_{\beta \rightarrow \infty} e^{-\beta H_A},
\eeq
we see that in this case we have a very sharp notion of entanglement thermodynamics where the temperature $T$ is uniform and zero.

Now suppose that $\rho_A$ violates the area law.  Then it must be the case that $G(H_A) \geq e^{S(A)}$, so $H_A$ has an enormous ground state degeneracy which grows with system size.  Indeed, \cite{2012PhRvL.109t7202B} showed precisely this fact for the model with $\log(L)$ entropy. Hence we see that these models must have peculiar thermodynamic properties with many low-lying states.  This is a precise sense in which the frustration-free models deviate from the ``reasonable" criteria of Ref.~\cite{2013Swingle-Senthil}. We hasten to add that this in no way undermines the interestingness of the frustration-free models; we simply understand better now how these models differ from the familiar examples of scale invariant states. CFTs on an interval, for example, do not have a large ground state degeneracy. In particular,
in the models of \cite{2012PhRvL.109t7202B,2014arXiv1408.1657M},
the thermal to entanglement crossover function discussed in Ref.~\cite{2013Swingle-Senthil} must take an unusual form.
\footnote{We note that \cite{2014arXiv1408.1657M} describes a modification
of these models which is not frustration-free
but exhibits similar entanglement phenomena; this too
must have a large density of low-lying states.}

\textit{Relation to s-source theory.} The above considerations show that the thermodynamic exponent $\theta$ plays a crucial role in the ground state entanglement properties. This connection can be clearly displayed by appealing to the notion of $s$ source RG fixed points defined in \cite{2014arXiv1407.8203}. The idea is to classify ground states based on how much entanglement is necessary to create them. Briefly, a phase is an $s$ source fixed point if $s$ copies of the ground state at linear size $L$ are necessary to produce one copy of the ground state at linear size $2L$.

All $s$ source fixed points obey the entropy bound $S(2R) \leq s S(R) + k R^{d-1}$, and we generally expect the bound to be saturated. Assuming the bound is saturated, we can apply the entropy recursion formula to the half-space region discussed above assuming that the ground state is entangled at all
length scales shorter than $w$ (the IR cutoff, a correlation length). The entropy is then
\bea
S(R) && = k \frac{R^{d-1}}{a^{d-1}} \sum_{n=0}^{\log(w/a)} \frac{s^n}{(2^{d-1})^n} \nonumber \cr \\
&& = k \frac{R^{d-1}}{a^{d-1}} \left( \frac{1 - (s/2^{d-1})^{\log(w/a)}}{1 - (s/2^{d-1})}\right).
\eea
Although the coefficients may not be correctly reproduced, this expression should correctly predict the scaling structure --  including the scaling structure of subleading terms. Taking the logarithm base two and writing $s = 2^{\log(s)}$ the subleading term scales like
\beq
S_{\text{sub}} \sim \frac{R^{d-1}}{a^{d-1}} \frac{a^{d-1-\log(s)}}{w^{d-1-\log(s)}}.
\eeq

Going back to Eq.~\ref{eq:entestimate}, the local thermodynamic calculation also predicts a subleading term of the form $\frac{R^{d-1}}{w^{d-1-\theta}}$.
Although Eq.~\ref{eq:entestimate}
applies to the maximum entropy locally consistent state $\sigma_A$, the
scaling structure of subleading terms in the entropy should be identical to those in the entropy of the actual subsystem state $\rho_A$. This claim amounts to assuming that there is no phase transition encountered in going from $\rho_A$ to $\sigma_A$ to $\sigma_A(\tau)$ (where the local approximation is justified). Demanding that the subleading terms match gives $\log(s) = \theta$ or
\beq
s = 2^{\theta}.
\eeq
This is a strong result which establishes an intimate connection between thermodynamic and entanglement properties. It should be noted, however, that this result may not apply to frustration-free Hamiltonians since their local entanglement temperature is $T=0$. It also remains to construct the unitary transformation performing the mapping from $L$ to $2L$; we have only shown that the entanglement scaling is consistent with being an $s$ source fixed point with $s=2^\theta$.
Examples of explicit RG circuits will be provided in \cite{sqrt-draft}.

\textit{Discussion.} We gave an argument for the area law in a wide class of scale invariant phases of matter. The argument worked by mapping to the problem of bounding the entanglement entropy to a problem of estimating the entropy of a local thermodynamic state. The key piece of thermodynamic data is the hyperscaling violation exponent $\theta$, while the dynamical exponent $z$ played little role. This result firmly establishes the intuition that highly entangled states of matter must have many low lying spatially extended excitations. We also discussed the analogous statement for the special class of frustration-free Hamiltonians and showed that they must possess an anomalously large number of low lying excitations in order to violate the area law. Finally, we related our considerations to the recently introduced $s$ source theory and provided another way to quantify the amount of entanglement in the ground state by relating the parameter $s$ to $\theta$.

The formalism accounts for the entanglement properties of all experimentally-realized states of matter
known to us which are stable and reach equilibrium. It also accounts for the entanglement properties of several infinite families of candidate states (like conformal field theories in general and critical Fermi surfaces) that might describe as yet mysterious experimental systems. Requiring stability seems essential since one can concoct fine-tuned or pathological Hamiltonians with highly entangled ground states (the frustration-free case likely being an example of this). On the other hand, glassy physics and other examples of failures to come to equilibrium are ruled out not because the present formalism necessarily fails (although it might), but because those cases deserve a separate and careful exposition. Among the states of matter that do fall into the present discussion, a number of interesting examples are displayed in Table 1. The point of this (not exhaustive) table is to emphasize that a great many experimentally realized quantum states of matter are covered by the gapped analysis in \cite{2014arXiv1407.8203} or the present gapless analysis and plausibly fit within the $s$ source theory.
\begin{table}
\begin{center}
\caption{Notation: SSB = spontaneous symmetry breaking, I/F QHE = integer/fractional quantum Hall effect, QCP = quantum critical point, QED = quantum electrodynamics, QCD = quantum chromodynamics. QCD is listed as $s=1^*$ because although the theory is ultimately gapped, the correlation length diverges in units of a hypothetical short-distance cutoff length which is taken to zero. Some representative calculations and further references can be found in \cite{Callan:1994py, Eisert:2008ur, PhysRevLett.96.010404, PhysRevLett.96.100503, Swingle:2009bf, Swingle:2010yi, Calabrese:2009qy, 2012PhRvL.108l0401C, 2011arXiv1112.5166M, Grover:2011fa, 2013NJPh...15b5002G, 2014arXiv1408.1094P, 2011PhRvL.107f7202Z, 2014arXiv1407.8203}.}
\begin{tabular}{|p{4cm}|p{1cm}|p{.7cm}|p{.7cm}|p{1.5cm}|}
  \hline
  State of matter & $z$ & $s$ & $\theta$ & EE \\
  \hline
  Insulators, etc. & Gap & $0$ & n/a & Area \\
  \hline
  SSB, discrete & Gap & $0$ & n/a & Area \\
  \hline
  IQHE (invertible) & Gap & $1$ & n/a & Area \\
  \hline
  FQHE & Gap & $1$ & n/a & Area \\
  \hline
  Topological states & Gap & $1$ & n/a & Area \\
  \hline
  SSB, continuous ($d>1$)& $1$ & $1$ & $0$ & Area \\
  \hline
  QCP (conformal), $d=1$ & $1$ & $1$ & $0$ & Area*Log \\
  \hline
  QCP (conformal), $d>1$ & $1$ & $1$ & $0$ & Area \\
  \hline
  Quadratic band touching & $2$ & $\leq 1$ & $0$ & Area \\
  \hline
  Fermi liquids & $1$ & $2^{d-1}$ & $d-1$ & Area*Log \\
  \hline
  Spinon Fermi surface & $3/2$? & $2^{d-1}$ & $d-1$ & Area*Log \\
  \hline
  Diffusive metal, $d=3$ & $2$ & $2^{d-2}$ & $d-2$ & Area \\
  \hline
  QED & $1$ & $1$ & $0$ & Area \\
  \hline
  QCD & Gap & $1^*$ & $0$ & Area \\
  \hline
\end{tabular}
\end{center}
\end{table}

To elaborate on an interesting case, \cite{2014arXiv1408.1094P} recently argued convincingly that the entanglement entropy of a diffusive metal (meaning a metal in the presence of static disorder with a diffusion pole in the density-density correlator) obeys the area law, unlike its clean cousin, the Fermi liquid. This is visible in the present framework as follows. The low energy density of states is not strongly modified by disorder, so the thermal entropy still scales like $s \sim T$. However, the metal is now diffusive rather than ballistic, so energy scales with wavenumber like $\omega \sim k^2$ and hence $z=2$. Requiring $s \sim T^{(d-\theta)/z} \sim T$ forces $\theta = d-2$ which is below the area law violation threshold.

A more complicated application of the formalism is provided by the exciton Bose liquid state of \cite{PhysRevB.66.054526}. This state of matter defined in $d=2$ has a peculiar dispersion relation which resembles $\omega^2 \sim k_x^2 k_y^2 $ near $k=0$. This dispersion relation naively suggests $z=2$ and $\theta=0$, but this is only part of the story. Because the dispersion is zero for all $k_x$ when $k_y = 0$ (and vice versa) the state also possesses lines of zero energy states. The $k_y =0$ line has $z=1$ with a variable velocity depending on $k_x$ and effectively has $\theta=1$. Thus although the thermal entropy goes like $s \sim T\log(1/T)$ the entanglement is dominated by the $\theta=1$ zero energy lines and yields a simple logarithmic violation. This violation has been seen explicitly in \cite{2013PhRvL.111u0402L} which realizes the counting argument of \cite{Swingle:2010yi}.

Hyperscaling violation necessitates the presence of an extra length scale $\ell$ in the density of states,
for example in order that the entropy density have units of length${}^{-d}$:
$s(T) = T^{{ d- \theta \over z}} \ell^{-\theta} $.
In the case of metals, this length scale is the Fermi wavelength.
We have assumed in various places, in particular in claiming that
in the half-plane geometry $ \xi = x$, and that this microscopic length scale does not enter.
This assumption is physically sensible and accords with all examples we are aware of; a further general argument is presented in an appendix.

The derivative expansion was carefully justified by introducing the fictitious temperature parameter $\tau$ in $\sigma_A(\tau)$. This trick gives most directly a bound on the entropy of the state of interest, but unless a phase transition is encountered as a function of $\tau$, one concludes that the entropy of $\sigma_A(\tau=1)$ has the same scaling structure as the entropy of $\sigma_A(\tau \gg 1)$. Similarly, although the entropy of $\sigma_A$ is only guaranteed to bound the entropy of $\rho_A$, one again expects the entropies to share the same scaling structure. This is because if $A$ were the entire system then $\sigma_A$ would equal $\rho_A$, so we expect that as $A$ is made larger the local approximation captures more and more of the entropy of $\rho_A$. This can be explicitly checked in some cases, e.g.~in Lorentz-invariant systems because $-\log(\rho_A)$ is local near the boundary of $A$ for any region $A$ (and this is where most of the entropy arises according to the thermodynamic argument).

The framework developed here, besides yielding a strong argument for the area law in scale invariant quantum states of matter, is of broader interest. Similar arguments have been used successfully for gapped phases of matter \cite{2014arXiv1407.8203} (there the analog of $\theta$ turns out to be the ground state degeneracy). It would be interesting to develop further the idea of entanglement thermodynamics into a full fledged theory of entanglement hydrodynamics, e.g.~in dynamical settings. For example, does the state $\sigma$ obey a simple dynamical equation? If so, this would be interesting since this equation would naturally include dissipation as information is lost into the exponential complexity of the quantum state.

\vskip.2in
{\bf Acknowledgements.}
This work was supported in part by
funds provided by the U.S. Department of Energy
(D.O.E.) under cooperative research agreement DE-FG0205ER41360,
in part by the Alfred P. Sloan Foundation. BGS is supported by funds from FQXi and the Simons Foundation.
\bibliographystyle{ucsd}

\bibliography{rg_area_law}

\appendix

\section{Further results on the local approximation}

Consider a translation invariant system with thermal entropy density $s(T)$. The total entropy of a volume $V$ at temperature $T$ is thus $V s(T)$. Now consider the same system with a position-dependent temperature, so that the state of the system is
\beq
\rho(\{T(x)\}) = \frac{\exp\left(- \sum_x \frac{H_x}{T(x)}\right)}{Z(\{T(x)\})}.
\eeq
Suppose now that the temperature $T(x)$ is approximately constant, $T(x) = T + \delta T(x)$ with $\delta T(x)\ll T$. Then the state of the system is close to the translation invariant thermal state. The first order correction is
\beq
\delta \rho = \rho(\{T(x)\}) - \rho(T) \approx \left(\sum_x \frac{(H_x - \langle H_x\rangle) \delta T(x)}{T^2}\right) \rho(T).
\eeq
We have not been careful about the operator ordering, but we will only use this expression within a trace so no harm can arise to first order in $\delta T/T$.

The change in the entropy is
\beq
\delta S = S(\rho(\{T(x)\}) - S(\rho(T)) \approx - \text{tr}\left(\delta \rho \log(\rho)  \right).
\eeq
This expression reduces to a two-point function of $H_x$,
\beq
\delta S \approx \frac{1}{T^3} \sum_{x,y} (\langle H_x H_y \rangle - \langle H_x \rangle \langle H_y \rangle ) \delta T(x).
\eeq
Translation invariance in the uniform temperature state implies that the connected two-point function $\langle H_x H_y\rangle - \langle H_x \rangle \langle H_y \rangle$ depends only on $x-y$. Thus after the sum over $y$ is performed, the change in the entropy is simply
\beq
\delta S = \frac{1}{V T^3} (\langle H^2\rangle - \langle H\rangle^2) \sum_x \delta T(x)
\eeq
where $H = \sum_x H_x$.

Now compare this expression with the expression from local thermodynamics,
\beq
\delta S \approx \sum_{x \in V} s(T(x)) - V s(T) \approx \sum_x \partial_T s(T) \delta T(x).
\eeq
$T \partial_T s(T) = c(T) $ is the heat capacity,
which is in turn related to energy fluctuations by
\beq
c(T) = \frac{1}{V T^2} (\langle H^2\rangle - \langle H\rangle^2).
\eeq
We conclude that to first order
\beq
S(\rho(\{T(x)\})) = \sum_x s(T(x))
\eeq
where $s(T)$ is the translation invariant entropy density.

We wish to argue for a stronger statement, namely that
\beq
S(\rho(\{T(x)\})) \approx \sum_x s(T(x))
\eeq
whenever $T(x)$ varies slowly on the scale of the local correlation length. This is the content of the derivative expansion discussed in the main paper. To quantify the validity of the derivative approximation it is useful to look at the derivatives of the entropy with respect to $T(x)$.  It will be useful to distinguish two averages, $\langle ... \rangle$ over the full non-uniform state,
and $\langle ... \rangle_{T(x)}$ over the uniform state with uniform temperature $T(x)$.

Computing $f_1 = \partial_{T(x)} S(\rho(\{T(x)\}))$ gives
\beq
f_1 = \left\langle \frac{H_x - \langle H_x\rangle}{T(x)^2} \sum_y \frac{H_y}{T(y)}\right\rangle = \sum_y \frac{\langle H_x H_y\rangle - \langle H_x \rangle \langle H_y \rangle}{T(x)^2 T(y)}.
\eeq
Similarly, computing $f_2 = \partial_{T(x)} \sum_y s(T(y))$ gives
\beq
f_2 = \frac{c(T(x))}{T(x)} = \sum_y \frac{\langle H_x H_y \rangle_{T(x)} - \langle H_x \rangle_{T(x)} \langle H_y \rangle_{T(x)}}{T(x)^3}.
\eeq
Comparing these two expressions gives a quantitative condition. The connected two-point function of the Hamiltonian ``density" must be approximately equal in the two states. Assuming this connected correlation function is exponentially decaying (due to the finite correlation length) and also exponentially insensitive to distant properties of the system (again, the correlation length), one concludes that the entropy of the temperature dependent state is well approximated by the sum of the translation invariant entropy density evaluated at the local temperature.

\section{Further justification of effective temperature scaling}

On general scaling grounds the effective temperature in $\sigma_A$ was assumed to go like $T(x) \sim 1/x^z$. The only other possibility is $T=0$ which is realized in frustration-free systems and which we do not consider in the appendix. However, this argument is slightly subtle when $\theta \neq 0$ since dimensional analysis implies that at least one additional length, call it $\ell$, does not decouple in the low energy limit. Perhaps the effective temperature only scales like $T(x) \sim f(x/\ell)/x^z$ with $f$ an undetermined scaling function?

The first argument in favor of $f=1$ begins with the observation that the (momentum) dimension of the energy density is $d-\theta + z$. In the presence of an infrared length scale $x$ (like the distance to a boundary), the energy density should to be modified by an additive shift of the form $\delta e/x^{d-\theta+z}$. This is exactly what is obtained above when $T(x) \sim 1/x^z$.

The second argument doesn't directly constrain $f$ but says that either the entropy bound is good or translation invariance of the local data must be broken. If $f=1$ then the analysis above goes through and the entropy is bounded by an area law for $\theta < d-1$. Suppose now that $f$ goes to zero as $x \rightarrow \infty$. Then a quick calculation confirms that the derivative expansion will not be valid as the system is approaching the ground state too rapidly as $x \rightarrow \infty$. However, the effective temperature is much lower than in the case $f=1$, so the entropy of a state with decaying $f$ will be upper bounded by the entropy of a state with $f=1$ so that the entropy bound is still valid. Finally, suppose $f$ goes to infinity as $x \rightarrow \infty$. Then the derivative expansion will be arbitrarily good as $x$ increases, but then it will be possible to detect a failure of translation invariance in the local data. This is so because a correlation length much smaller than $x$ will imply that the boundary at $x=0$ is not observable and thus the expectation value of $H_x$ will depend on $x$.

The third and final argument proceeds by analyzing energy fluctuations and directly shows $f=1$ given the assumption that energy fluctuations in the maximum entropy state scale the same way as in the true state. This assumption is again made plausible by the fact that the true ground state is recovered when $A$ is the total system. The argument begins by observing that there are two ways to compute energy fluctuations, by directly integrating the connected energy-energy correlation function and by integrating the heat capacity over all space.

Let $\Delta_H$ denote the spatial scaling dimension of the energy density (this dimension can be different from $d-\theta+z$ which is better understood as a temporal scaling dimension). Then the connected 2-point function $\langle H_x H_y \rangle$ will decay like $|x-y|^{2 \Delta_H}$ and the energy fluctuations (fluctuations of $H_A$) of a strip will have a UV finite term going like
\beq
\Delta H_A^2 \sim \frac{L^{d-1}}{w^{2\Delta_H - d -1}}.
\eeq
$\Delta_H$ may be determined from ordinary thermodynamics by observing that energy fluctuations are given by $T^2 c(T)$ where $c(T) \sim T^{(d-\theta)/z}$ is the heat capacity. At temperature $T$ the energy fluctuations are obtained from the $k=0$ limit of the Fourier transform of the equal time energy-energy 2-point function. Scaling again determines that $\langle H H \rangle (k\rightarrow 0) \sim T^{(2 \Delta_H -d)/z}$ where the $-d$ is from the integral over space defining the Fourier transform. Setting the two expressions for the energy fluctuations equal gives $2\Delta_H =2 d + 2z -\theta$. Note a hidden assumption here, that all directions in space scale the same way; this assumption can be violated, e.g. in a 2d array of decoupled 1d wires, but a more general analysis can be performed in such cases leading to the same overall conclusion.

Turning now to the local thermodynamic calculation for the strip in the ground state, one sees that if and only if $T\sim 1/x^z$ will the energy fluctuations obtained by integrating $ T(x)^2 c(T(x))$ be equal to those obtained from the double integral of $\langle H_x H_y \rangle$ over $A$. Hence requiring that the scaling of energy fluctuations be reproduced forces $T(x) \sim 1/x^z$.

As a final note, $T(x)$ may also contain terms which oscillate at a wavelength set by $\ell$. Since $\ell$ is a microscopic length these oscillations are far too rapid for the local approximation to be applicable on that scale. This is in fact why such oscillations are allowed (otherwise the local approximation would lead to badly non-translation invariant local data). If such oscillating terms are subleading compared to the dominant scaling of $T(x)$ then all is fine. An envelope to the oscillation decaying slower than $1/x^z$ is again ruled out because it would lead to non-translation invariant local data.

\end{document}